\documentclass[preprint,12pt]{elsarticle}

\usepackage{amssymb}
\usepackage{lipsum}
\usepackage{url}

\usepackage{lineno}

\journal{Nuclear Instruments and Methods A}

\begin{document}

\begin{frontmatter}



\title{Measurements of the Birks' coefficient of GAGG:Ce using hard X-rays}


\author[first]{Merlin Kole}  
\author[second]{Nicolas De Angelis}  

\affiliation[first]{organization={University of New Hampshire, Space Science Center},
            city={Durham},
            postcode={03824}, 
            state={New Hampshire},
            country={USA}}

\affiliation[second]{organization={INAF-IAPS},
            addressline={via del Fosso del Cavaliere 100}, 
            city={Rome},
            postcode={00133}, 
            country={Italy}}

\begin{abstract}
Inorganic scintillators continue to be widely used within astrophysical X-ray and gamma-ray detectors. This is in part thanks to the development of new scintillators, such as GAGG:Ce, as well as the availability of new scintillator readout sensors such as Silicon Photomultipliers and Silicon Drift Detectors. In order to use such scintillator materials for spectrometry or polarimetry, a detailed understanding of their response is important. One parameter that can affect the scintillator performance, particularly at lower photon energies, is their Birks' coefficient, which correlates the relative light yield to the ionization energy density. While for many high-Z inorganic scintillators this effect can be ignored, for GAGG:Ce this appears to not be the case. Here we provide a measurement of the Birks' coefficient for GAGG:Ce using data from a detector irradiated in the 20-80~keV energy range at the LARIX-A X-ray beam in Ferrara, Italy. While the effects due to Birks' law are visible below 30 keV, they also significantly influence the performance of GAGG:Ce performance near one of the K-edges, affecting both the measured gain and the energy resolution. Here, we use beam test data to derive the Birks' coefficient from GAGG:Ce. The results indicate that for usage in hard X-ray and soft gamma-ray missions, this coefficient has a significant effect on the measurements.
\end{abstract}




\end{frontmatter}




\section{Introduction}
\label{introduction}

Cerium-doped $\mathrm{Gd_3(Ga, Al)_5O_{12}}$ (GAGG:Ce) is an inorganic scintillator with many advantages in the field of X-ray and gamma-ray astrophysics. Its effective atomic number of $Z_{eff}=54.4$ and large density of $\rho=6.63\,\mathrm{g/cm^3}$ make it more efficient for stopping photons than other common scintillator materials like CsI or NaI. Although materials, such as BGO, exist with a higher stopping power, its high scintillation efficiency of around 45'000 optical photons/MeV ensures a good energy resolution. Furthermore, it has a relatively fast decay time of the order of $100\,\mathrm{ns}$ which, unlike materials like BGO and CsI, is very stable with temperature \cite{yoneyama_etal_2018, Kole2025_b}.  Its non-hygroscopic nature furthermore makes it easy to handle and to incorporate into detector designs. Based on these characteristics this relatively new scintillator type has been proposed for a range of both balloon-borne and space-based astrophysical missions over the last two decades. Some examples are GARI \citep{GARI}, GRID \citep{GRID} GTM \citep{GTM}, CXBe \cite{CXBe} and the POLAR-2 spectrometer (BSD) \citep{POLAR-2_paper, POLAR-2_paper2} which use it with a SiPM-based readout. The GRAPE \cite{GRAPE} mission has furthermore used it both for spectrometry and polarization measurements. In addition, the HERMES mission uses this material with an SDD readout \citep{campana_etal_2023}. It is also foreseen to be used in anti-coincidence detectors for both the balloon-borne ASCENT mission \cite{Kole2025_b} proposal as well as for the space-based HERD mission~\cite{HERD_ACS}.

Performing accurate spectrometry and polarization measurements requires an in-depth understanding of the scintillator materials used. For example, the light yield and its dependency on temperature are important factors to consider as they can alter the response of an instrument over its full energy range. Thanks to its popularity in high-energy astronomy missions, most of GAGG:Ce's parameters, such as its temperature dependence \citep{yoneyama_etal_2018, Kole2025_b}, are now well understood. At lower energies, some scintillators also experience quenching effects due to Birks' law. This law describes the quenching of the light yield in scintillator materials as a function of the ionization density. In an ideal detector, the amount of scintillation light produced is linearly proportional to the energy deposited. However, in some scintillators, particularly organic plastic ones, at high ionization densities, the linearity relation breaks down. When large amounts of energy are deposited in a small area, complex effects like saturation start to appear, resulting in lower efficiencies for high values of \( \frac{dE}{dx} \) \cite{Birks1951}.

Birks' law quantitatively captures this nonlinearity by relating the scintillation light output per unit path length, \( \frac{dL}{dx} \), to the stopping power, \( \frac{dE}{dx} \), of the ionizing particle. The empirical formula used to describe the relation is as follows:

\begin{equation}\label{eq:1}
    \frac{dL}{dx} = \frac{S \frac{dE}{dx}}{1 + kB \frac{dE}{dx}}
\end{equation}

where $S$ is the absolute scintillation efficiency, and $kB$ is Birks' coefficient, which characterizes the quenching behavior of the scintillator.

As, following the Bethe-Bloch equation, the $\frac{dE}{dx}$ is inversely proportional to the velocity of the charged particle, it increases rapidly for electrons in the keV energy range when these start to slow down \cite{Bethe}. Therefore, the Birks' effect results in lower scintillation efficiencies for low-energy X-rays. An example is the EJ-248M scintillator material, where the average energy measured from gamma-rays with energies below $\sim100\,\mathrm{keV}$ is significantly affected. This prompted detailed measurements for the POLAR mission of this material, resulting in a $kB$ of $0.143\,\mathrm{mm/MeV}$ \cite{ZHANG201594}. For inorganic scintillators, the Birks' coefficient is typically significantly lower. Examples of this are BGO with $kB=0.0038\,\mathrm{mm/MeV}$ \cite{BGO} and NaI with $kB=0.0046\,\mathrm{mm/MeV}$ \cite{BGO}.

While important for X-rays, the effect is even more important for charged particle measurements, especially heavy ions, where, as the ionization increases with the atomic number, the dependency of equation \ref{eq:1} leads to strong quenching effects. 

In this paper, we use photon beam data collected using a detector which contains GAGG:Ce crystals to derive the Birks' coefficient of this scintillator material. In section \ref{sec:GAGG}, we introduce the measured effects of Birks' law on the efficiency of GAGG:Ce around the K-edge. We then provide a brief overview of the measurements performed at the LARIX-A beam facility in section \ref{sec:Measurements}. In addition, this section includes a discussion on other potential sources of non-linearity in the system and how these are mitigated. This is followed by an overview of the simulations used to reproduce these results in section \ref{sec:sims} and a discussion of the derivation of the Birks' coefficient in section \ref{sec:analysis}, followed by conclusions.

\section{Non-linearity of GAGG:Ce at soft gamma-ray energies }\label{sec:GAGG}

\begin{figure}[h!]
\centering
 \includegraphics[height=.4\textwidth]{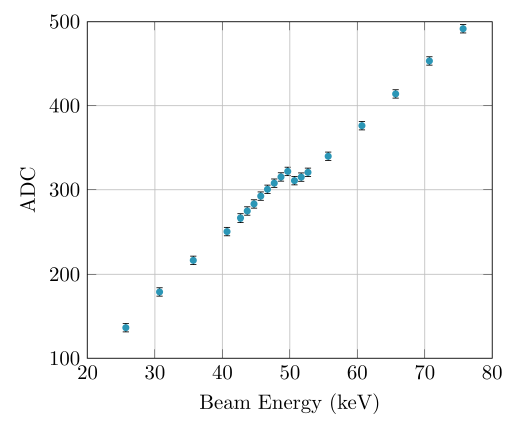}\includegraphics[height=.40\textwidth]{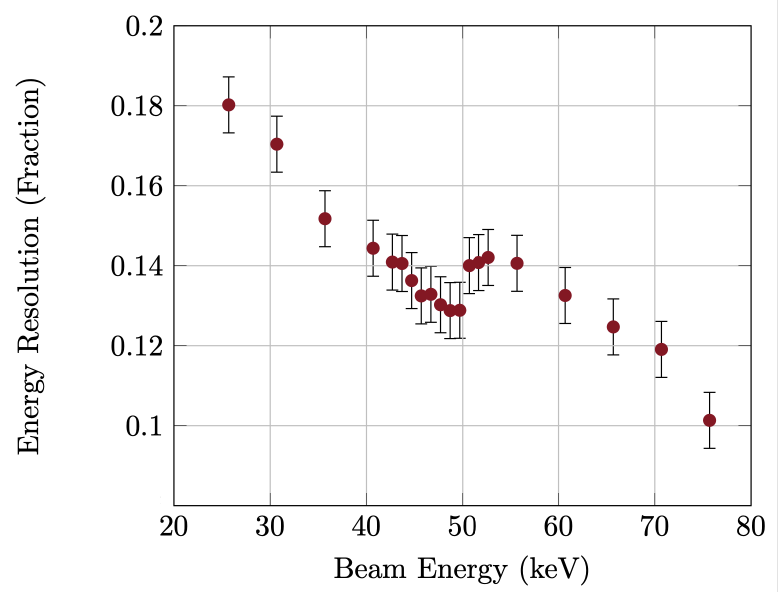}
 \caption{\textbf{Left:} The mean pulse height versus beam energy as measured by the POLAR-2 spectrometer prototype. A clear jump between 50 and 51 keV can be observed. \textbf{Right:} The energy resolution from the same detector as a function of beam energy. Here, an increase in the energy resolution can be observed at the same energy. }
 \label{fig:ADC_POLAR}
\end{figure}

While the non-linear behavior of GAGG:Ce at photon energies below $100\,\mathrm{keV}$ has been noted \citep{Kole2025_a,campana_etal_2023}, measurements of its Birks' coefficient using photon beams have not been performed yet. This is in part because the effect is not significant for energies exceeding $\sim30\mathrm{keV}$ where most spectrometers operate. An exception can, however, be found around the K-edge of Gd at $50.2\,\mathrm{keV}$. A similar behavior has been observed around the various binding energies in NaI \cite{NaI}. For GAGG:Ce this behavior was previously reported by \cite{campana_etal_2023}, and was measured using a prototype of the spectrometer of the POLAR-2 mission \cite{Kole2025_a} as well as the GRAPE detector \cite{OnateMelecio2025grape}. Figure \ref{fig:ADC_POLAR} shows the mean measured pulse height (in ADC) as a function of the beam energy along with the energy resolution as measured using the POLAR-2 prototype. This prototype, described in detail in \cite{Kole2025_a}, uses a Citiroc 1A ASIC along with a 12 bit analog to digital converter to produce the ADC spectra presented in this study.

The POLAR-2 prototype detector used for these measurements consists of an $8\times8$ array of GAGG:Ce crystals coupled to a SiPM readout. Details of this detector can be found in \cite{Kole2025_a}. The GAGG:Ce crystals have dimensions of $6\times6\times15\,\mathrm{mm^3}$. 

The sharp drop in the pulse height above the K-edge of gadolinium can be understood from Birks' law. While for energies below the K-edge the majority of the hard X-rays will be absorbed in a single photo-absorption interaction, above this energy a significant number will distribute their energy among two separate electrons. Above this energy, a photon can lose $50.2\,\mathrm{keV}$ through absorption by the K-edge electron while the remainder of its energy can be absorbed by a secondary electron. This secondary electron has the remaining energy, which for photons with energies just above $50.2\,\mathrm{keV}$ corresponds to only a few keV. This results in a high average $\frac{dE}{dx}$ and therefore a low scintillation efficiency, resulting in a lower total pulse height. In addition, as the energy of the incoming photon is, on average, more often spread between 2 electrons, the energy resolution deteriorates. 

\begin{figure}[h!]
\centering
 \includegraphics[width=9 cm]{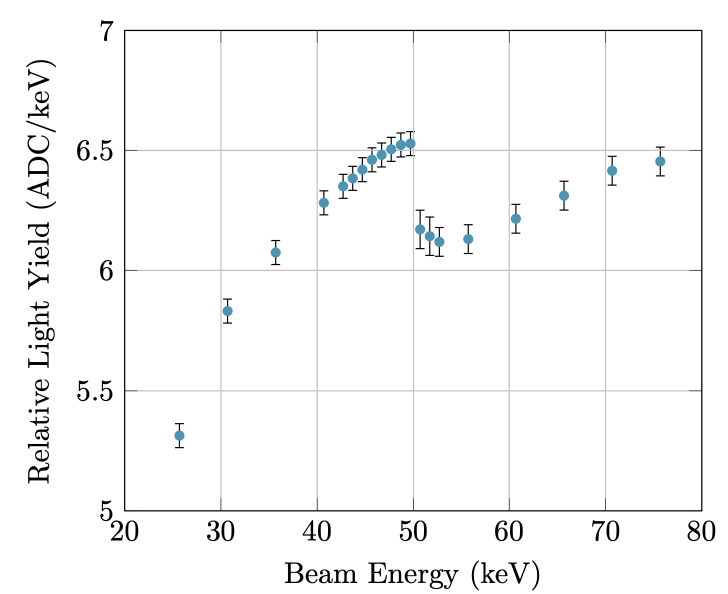}
 \caption{The mean pulse height divided by the beam energy versus the beam energy. The clear drop in efficiency of the scintillator above the K-edge remains visible, while non-linear effects at low energies also become visible.}
 \label{fig:ratio_POLAR}
\end{figure}

The overall effect of Birks' law on the GAGG:Ce measurements becomes clearer when plotting the ratio of the pulse height over the incoming photon beam energy as shown in Figure \ref{fig:ratio_POLAR}.

It should be noted here that not all photons with energies exceeding $50.2\,\mathrm{keV}$ will always produce 2 electrons. The probability of this is complex to calculate, as it also depends on the probability for the photons to escape the crystal after liberating the K-edge electron. This in turn depends on the interaction depth and the overall geometry of the crystal. A proper estimation therefore requires detailed Monte-Carlo simulations.

\section{Measurement Setup}\label{sec:Measurements}

\subsection{Detector Setup}

\begin{figure}[h!]
\centering
 \includegraphics[height=.30\textwidth]{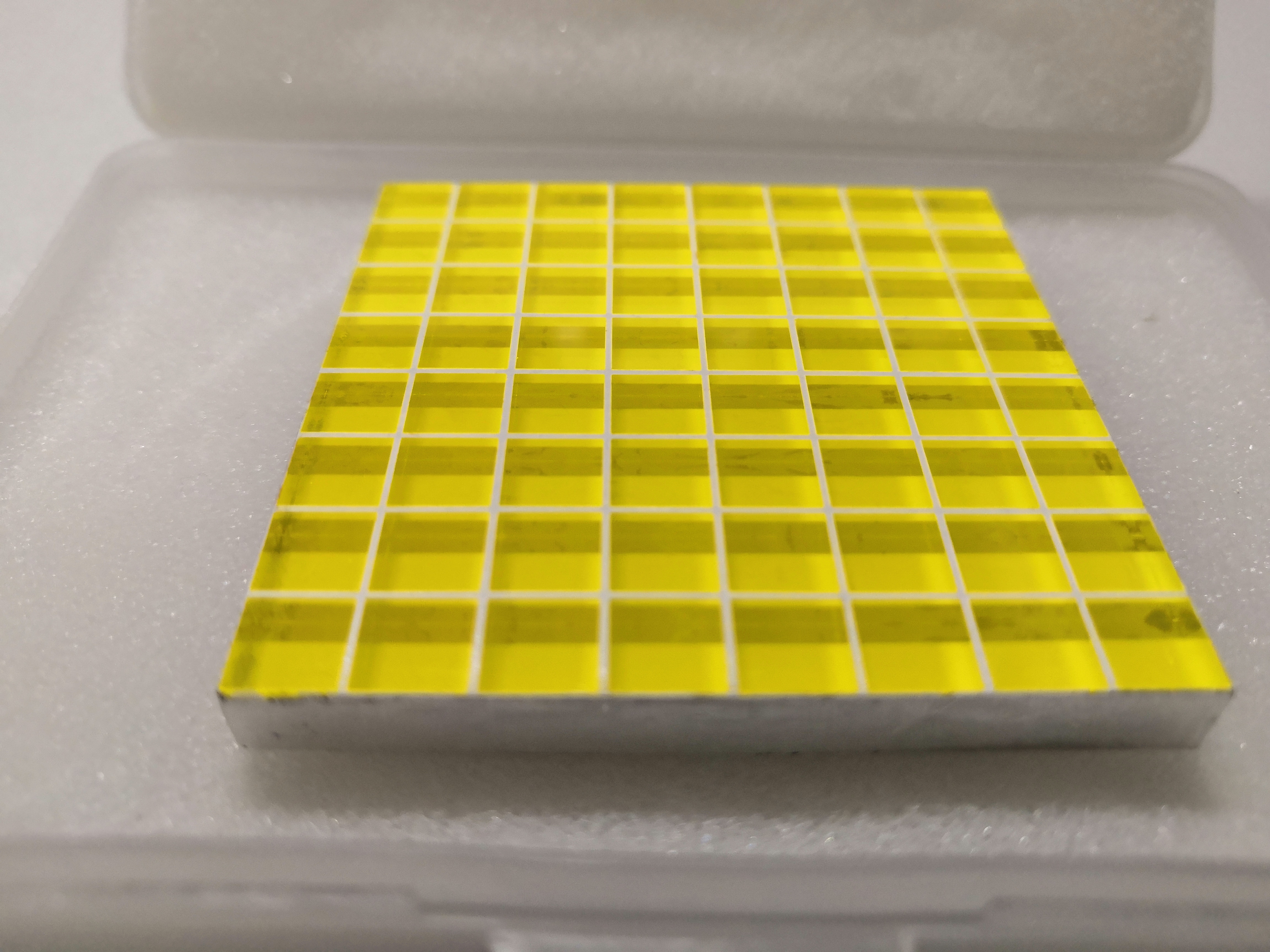}
  \includegraphics[height=.30\textwidth]{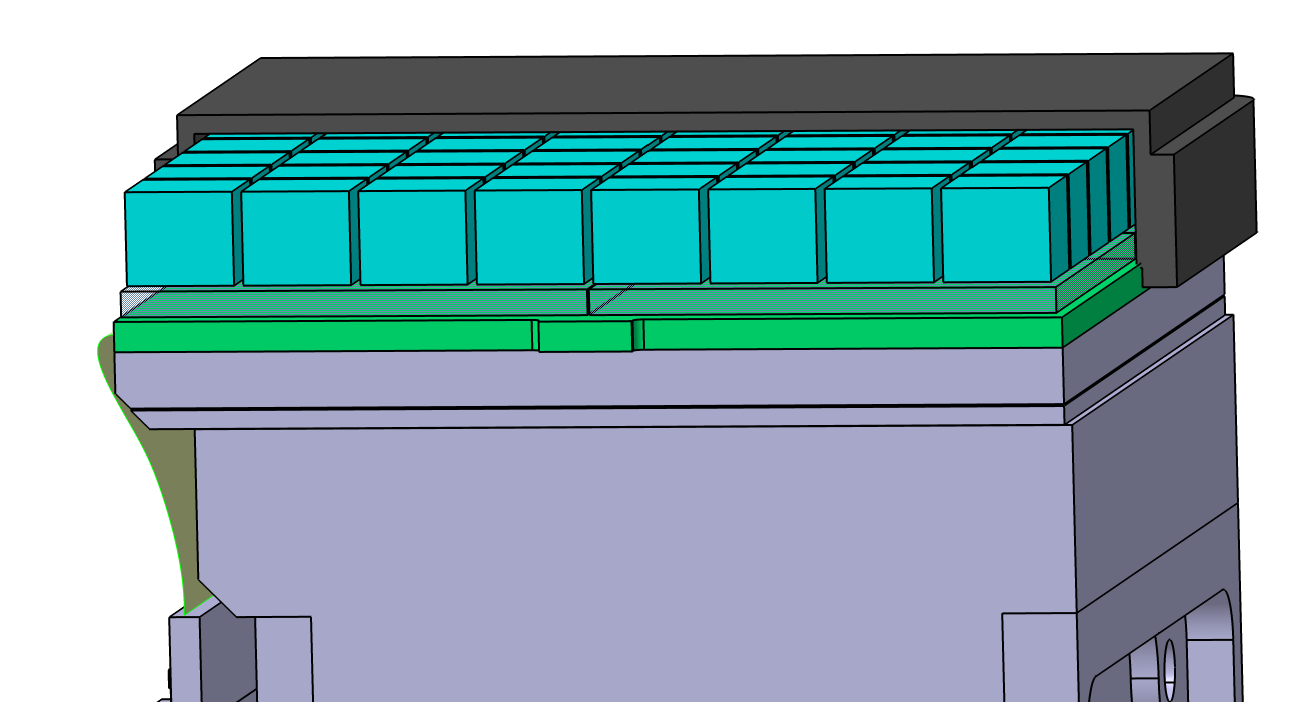}
 \caption{\textbf{Left:}The array of $8\times8$ GAGG:Ce crystals placed in the $\mathrm{BaSO}_4$ mechanics as used during this test. Taken from \cite{Kole2025_a}. \textbf{Right:} The schematic design of the full detector. The GAGG crystals are shown as teal blocks placed below a 1~mm-thick carbon fiber mechanical frame. The thin $\mathrm{BaSO}_4$ housing is now shown in this schematic. The SiPM array is shown in green below the crystals, with below it the readout PCB and further aluminum mechanics in gray.}
 \label{fig:GAGG_array}
\end{figure}

Measurements were first performed using an $8\times8$ array of GAGG:Ce crystals readout with a Citiroc 1A-based readout system originally developed for the POLAR-2 mission. This system is described in detail in \cite{Kole2025_a}. The GAGG crystals have dimensions of $6\times6\times15\,\mathrm{mm^3}$ and are placed in a mechanical structure made of $\mathrm{BaSO}_4$. Here the $\mathrm{BaSO}_4$ serves both as an optical decoupler between the different crystals while its high reflectivity also ensures a high light yield. The array can be seen in Figure \ref{fig:GAGG_array}.

Each GAGG:Ce crystal was coupled directly to a channel of a S13361-6075PE-04 SiPM by Hamamatsu \cite{SiPM}. The 64 SiPMs are in turn read out using 2 Citiroc 1A ASICs, which provide both a high and a low gain output both of which are digitized on the readout board. For the work presented here the low gain output was used. The overvoltage provided to the SiPMs was set to $3\,\mathrm{V}$ and was kept stable throughout the measurements. 

The system was placed on the XYZ table provided by the beam facility, allowing one of the crystals to be aligned to the beam. The alignment took place over a period of approximately 1 hour, during which the temperatures in the detector stabilized.  

\subsection{X-ray Beam Setup}

\begin{figure}[h!]
\centering
 \includegraphics[width=9 cm]{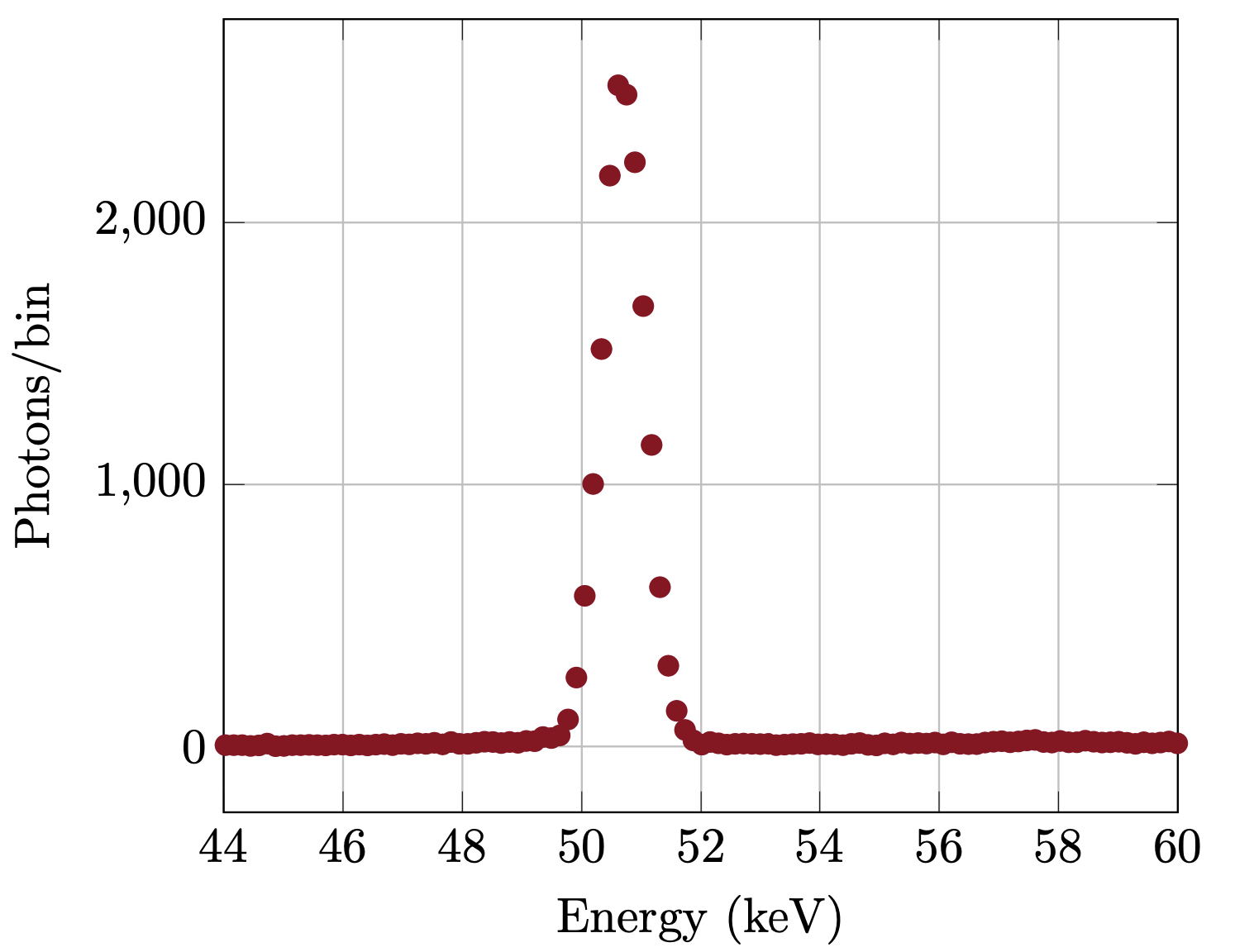}
 \caption{The energy spectrum from the beam as measured using the HPGe detector for a beam energy of 50~keV. A shift of 0.7~keV in the mean can be observed while the energy resolution corresponds to $0.75\%$.}
 \label{fig:HPGE_spec}
\end{figure}

The irradiation took place at the LARge Italian X-ray facility (LARIX-A) in Ferrara, Italy. This facility provides a 12~m long beamline with a Bosello X-ray tube coupled with a fixed-exit Bragg-Bragg monochromator. The setup can be tuned to produce a mono-energetic beam in the range of 10 keV to 200 keV. For the work here a beam with energies in the 25 to 75~keV range was used. Below 30~keV, the intensity of the beam starts to drop significantly \cite{campana_etal_2023}, and dead materials covering the crystals also start to become important. The energy spectrum of the beam was measured at several energies using an ORTEC nitrogen-cooled HPGe spectrometer with a beryllium entrance window provided by the facility. An example of the measured spectrum for a 50~keV beam is shown in Figure \ref{fig:HPGE_spec}. A small observed shift of 0.7~keV can be observed which was also found at beam energies 100~keV. Through discussions with the beam operators this was understood to result from a misalignment of the monochromator. While such a misalignment typically causes an offset in the energy which increases linearly with the beam energy, a correction for this had already been included by the beam operator. As a result a residual offset, which can vary with energy, remains while the larger linear correlation is removed. The energy resolution is consistent with that reported for the same facility in \citep{campana_etal_2023}. For the studies here, we therefore take the detailed energy resolutions as reported in \citep{campana_etal_2023} when simulating the beam. In addition, in order to be able to compare the measurements more accurately to simulations, the latter were performed including the 0.7~keV shift. It should be noted that we cannot confirm that the offset was consistently at 0.7~keV for all beam energies as it was only measured at 50~keV and 100~keV. We therefore performed additional studies where the offset was varied within 0.5~keV and found this resulted in a systematic error of several $\%$ to our final result as will be discussed towards the end of this paper. 

The beam used for these measurements was collimated using a set of adjustable tungsten plates placed at the end of the beam. The beam size used here was approximately $7\times7\mathrm{mm^2}$, thereby ensuring it covers a full crystal. The alignment of the detectors with the beam could be adjusted using an XYZ table. The table was adjusted such that the beam was pointed towards the center of the crystals with a precision of $\sim1\,\mathrm{mm}$. 

\begin{figure}[h!]
\centering
 \includegraphics[width=9 cm]{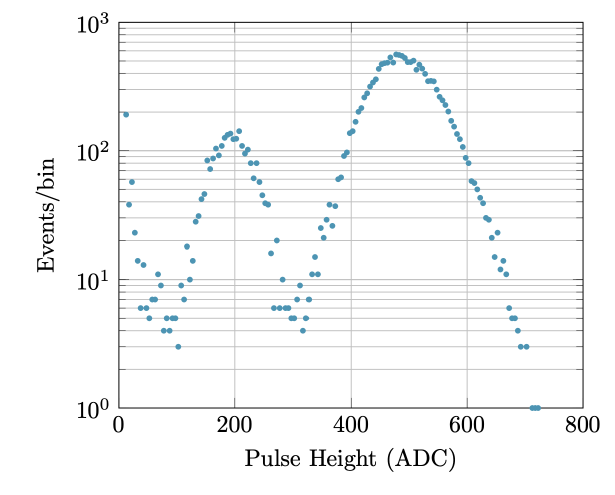}
 \caption{An example of the pulse height spectrum in ADC taken with a beam energy of 75~keV. The noise region can be observed at low ADC values along with 2 photo-peaks. The higher of these two corresponds to the 75 keV photo-peak while the lower is a 32 keV peak induced by the $\mathrm{BaSO}_4$ mechanical housing. }
 \label{fig:spec_example}
\end{figure}

A typical spectrum as measured using the setup with the 75~keV energy beam is shown in Figure \ref{fig:spec_example}. At low ADC values, the noise region can be observed. This is produced as a result of photons triggering some of the neighboring 63 channels. Despite the energy deposition in the studied channel being below its trigger threshold, a trigger in a neighbor results in the ADC value of this channel to be read out, leading, typically, to a noise value. For the work presented here, these events can be ignored. At higher ADC values, two photopeaks can be observed. The highest of these is the expected 75~keV peak, the second is at 32~keV and is induced by the K-edge emission of barium from the crystal's mechanical $\mathrm{BaSO_4}$ housing. This peak at 32~keV is consistently observed in all spectra with beam energies of ~40 keV and above. Although other sources could produce a peak around 32~keV, such as an escape peak from Gd when irradiated using 75~keV photons, the constant position of the peak indicates it results from a material outside of the GAGG:Ce crystal. Simulations of the full setup, discussed in \cite{Kole2025_a} confirmed the origin to be from the $\mathrm{K\alpha1}$ and $\mathrm{K\alpha2}$ lines from the $\mathrm{BaSO_4}$ housing. The choice of using $\mathrm{BaSO_4}$ for the mechanical housing was originally based on its optical properties which allow to construct the detector array without the need for additional reflective materials. Based on the measurements performed here the design has since been revised for the final space based detector.

To produce the pulse height versus beam energy plot presented in Figure \ref{fig:ADC_POLAR}, the spectrum was fitted using two Gaussians. One to account for the 32~keV peak and one for the photo-peak induced directly by the beam. Although for the spectrum presented here the $32\,\mathrm{keV}$ peak does not influence the fit result of the photo-peak at $75\,\mathrm{keV}$, at energies where the two overlap it is important to take it into account. Although initially seen as a nuisance in this analysis, the $32\,\mathrm{keV}$ peak was used to study any time dependencies in the gain, for example those due to small changes in temperature. The position of the peak was found to vary by less than 4 ADC ((0.6~keV), within the typical fitting error, during the data runs while no systematic shift in time was observed. The peak therefore allows to exclude any temperature induced gain shifts from the analysis within $2\%$.

\section{Other sources of non-linearity}

\subsection{SiPM non-linearity}

Apart from a non-linear response of the scintillator with incoming beam energy, both the SiPM readout and the electronics could induce similar effects. 

SiPMs can show non-linear behavior with measured energy due to saturation effects. The signal from a SiPM is produced when an optical photon initiates an avalanche or Geiger breakdown in the microcells which make up the sensor. While one Geiger breakdown can be observed as a single photo-electron peak, the signals observed in this study consist of the sum of tens of these induced by the scintillation light. While the number of microcells is large, it is finite, and only one breakdown can occur per microcell per energy deposition in the scintillator, after which it needs to recover. As the number of optical photons increases, the probability for a photon to reach a microcell that is already activated by another optical photon increases. Therefore, saturation effects start to become important at higher energies. Typically, non-linear effects start to become visible when the number of optical photons exceeds 2/3 of the number of microcells \cite{SiPM_sat} which for the SiPM used here (Hamamatsu S13361-6075) is 6400. This would therefore correspond to approximately 4000 optical photons. Below this number, the behavior of a SiPM can be considered linear within $1\%$.

The number of Geiger breakdowns, or photo-electrons, per deposited keV can be calculated by measuring the ADC position of the photo-electron peak along with the ADC position of a photo-peak induced by the beam. This was done previously for this system in \cite{Kole2025_a} where a value of $2.3\,\mathrm{p.e./keV}$ was found. This implies that for the data used here, the maximum amount of photo-electrons corresponds to roughly 200 corresponding to only $5\%$ of the microcells being activated. Using the equations from \citep{SiPM_sat} and the characteristics of the S13361-6075 MPPC and a GAGG crystal with a light yield of 45'000 optical photons/MeV and a light collection efficiency of $20\%$ we see that deviation from a fully linear behavior stays within $1\%$ in the energy range studied here.

\subsection{Readout Electronics non-linearity}

\begin{figure}[h!]
\centering
 \includegraphics[width=14 cm]{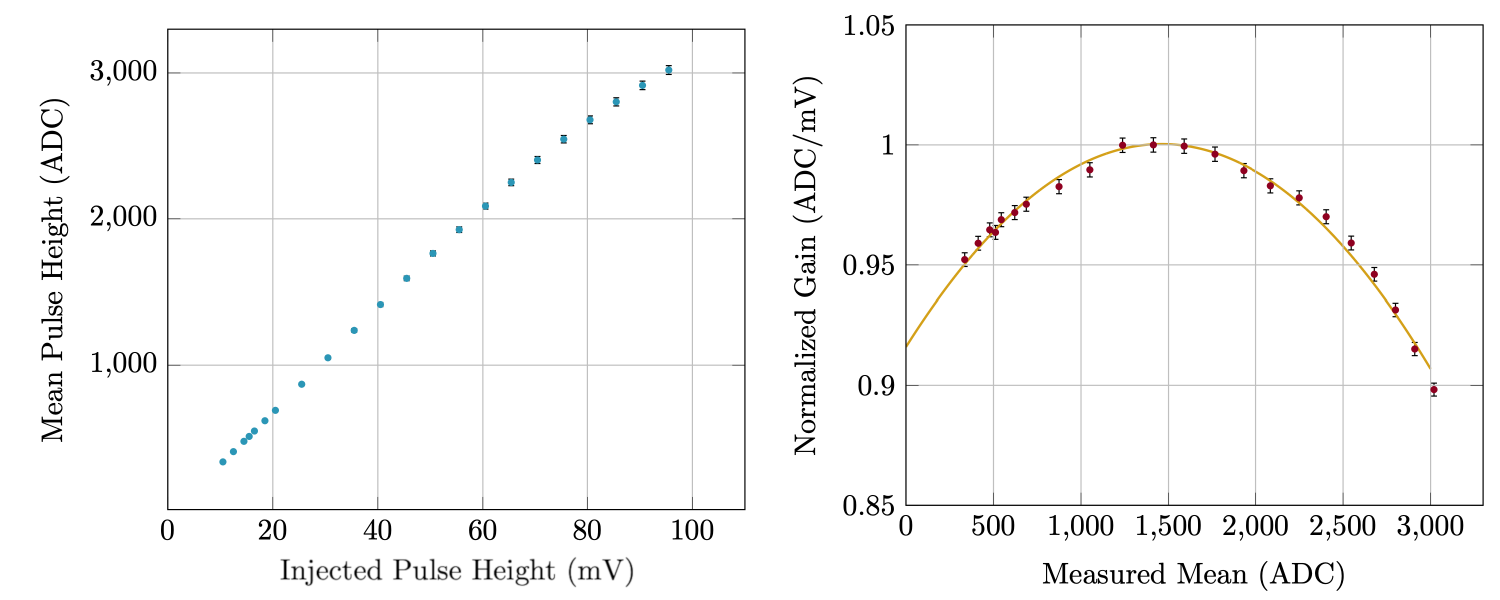}
  \includegraphics[width=7 cm]{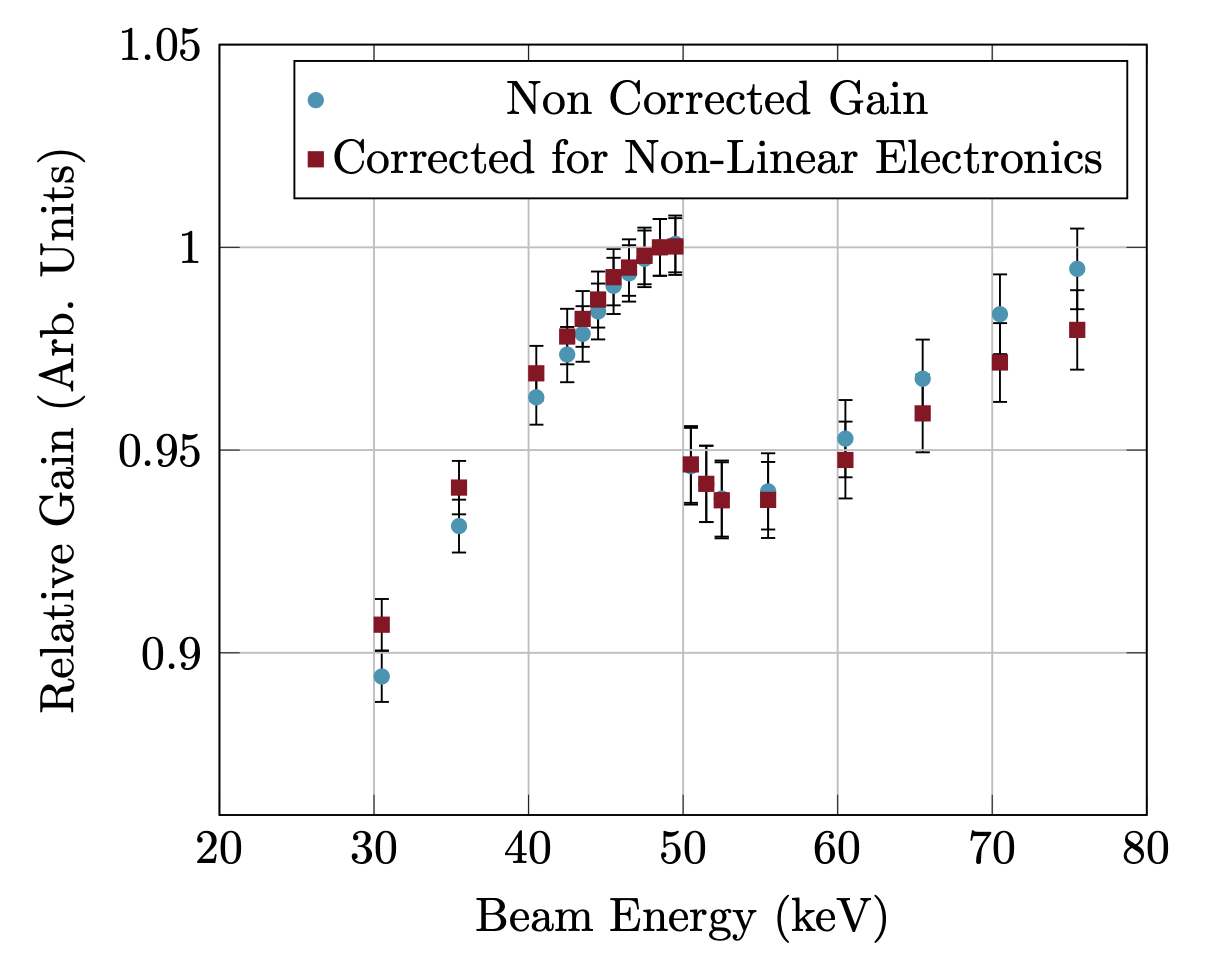}
 \caption{\textbf{Top Left:} The measured mean ADC value as a function of the injected charge as measured using the Citiroc-based POLAR-2 electronics. \textbf{Top Right:} The normalized gain, defined as the mean measured ADC value divided by injected charge, as a function of the mean measured ADC. The measured gain is fitted with a 2\textsuperscript{nd} order polynomial, resulting in a reduced $\chi^2$ of 0.8. \textbf{Bottom:} The relative gain as a function of energy both before (circles) and after correction (squares) for the non-linear electronic gain.}
 \label{fig:charge_lin}
\end{figure}

The linearity of the electronics can be tested by injecting pulses with known charges into the system. This was performed for the electronics used here using a pulse generator which injected signals directly into the ASICs. This allowed to test the full system independently from the effects of the SiPMs and scintillator. Pulses in the 10 to 100 mV range were injected in steps of 2 mV below 20 mV and 5 mV above this. This results in a series of normal distributions in the measured ADC spectrum. Each Gaussian was fitted, and the resulting mean can be plotted against the injected pulse height. The result is shown on the left of Figure \ref{fig:charge_lin}. Although the gain looks only to deviate from a linear one at high ADC values, deviations at lower ADC values become clear when plotting the gain instead. To do this, the measured mean ADC value was divided by the injected pulse height. This gain, normalized by the highest measured gain, was plotted against the measured ADC to produce the right side of Figure \ref{fig:charge_lin}. Here we see deviations from a perfectly linear system, which would result in a flat horizontal line, also at low ADC values.

The beam data used for this study was all contained in the 200 to 600~ADC range, where, although not severe, some non-linearity can be observed. In order to take this into account, the gain curve, now expressed in ADC such that it can be used for correcting beam data, from Figure \ref{fig:charge_lin} was fitted using a 2\textsuperscript{nd} order polynomial in the 0 to 3000~ADC range, where it had a reduced $\chi^2$ of 0.8, indicating it to be a good approximation. The fit result was used to correct the measured ADC spectra from the beam, resulting in shifts of the measured peak positions of several $\%$. The effects of this correction, which can be seen at the bottom of Figure \ref{fig:charge_lin}, are minimal, of the order of $1\,\mathrm{\mu m/MeV}$, on the final derived value of $kB$.

\section{Simulation Setup}\label{sec:sims}

The Geant4 framework \cite{Agostinelli2003GEANT4} was used to perform Monte Carlo simulations of the setup. In this setup, only the GAGG:Ce crystal was simulated along with the beam. Although other components in the measurement setup, such as the mechanical housing, can have an effect on the measured spectrum, they do not alter the position and width of the photo-peak and can therefore be ignored. The crystals were simulated with their actual dimensions along with the beam, which was simulated with a size of $7\times7\,\mathrm{mm^2}$ and a uniform intensity. It should be noted that the effects of the size of the beam and its alignment were studied by repeating the simulations with different settings for these variables. The beam size was varied by a factor of 2, and its position from the center was shifted up to values of 2 mm as well. No significant differences were observed when varying these values within their uncertainties.

The effects of Birks' law were taken into account using tools available for this within Geant4. Within Geant4, the Birks' coefficient (in mm/MeV) can be set for all the materials, and both the total deposited energy as well as the 'visible energy' can be extracted from the simulations. In these, the 'visible energy' corresponds to the deposited energy corrected for the Birks' coefficient. In Geant4, its effects are accounted for by applying the quenching factor, calculated using Eq. \ref{eq:1}, to each energy deposition produced through ionization. While the total deposited energy by an electron simply consists of all the energy it has lost through ionization, the visible energy includes the correction for the quenching along its ionization path.  

The simulations made use of the G4EmLivermorePhysics list, which is known to accurately model electromagnetic processes at the energy range of our interest. In addition, it is important to set up the simulations so that they are accurate for low energy depositions. To achieve this, the CutValue parameter, which sets the minimal range a secondary particle can travel in a material for it to be produced, was set to $0.0\,\mathrm{\mu m}$ for both the gamma and the e$^-$. This value ensures all secondary particles, regardless of their range, to be produced in the simulation. In addition, the step size used to simulate interactions was changed from its default value. This is of particular importance to this study as the measurement of $kB$ depends on the step size selected for $\frac{dE}{dx}$. The smaller the step size, the more accurate the results.

In Geant4 the step size limit $S_{\Delta}$ is calculated as:

\begin{equation}
    S_{\Delta} = \alpha R + \rho(1-\alpha)
\end{equation}

Here $R$ is the range of the particle and $\rho$ is a 'final range' parameter, a typical scale for the minimum step size with a default around 1 mm. Instead, $\alpha$ is a parameter
which determines the fraction of the remaining range of the particle covered in a calculation. When the remaining range of a particle is large compared to $\rho$, the step size is dominated by $\alpha R$, while, when it runs out of energy it becomes closer to $\rho$. As $\frac{dE}{dx}$ increases quickly for ionizing particles as they deposited their energy, it is important to have both $\rho$ and $\alpha$ here to be very small. This comes at the cost of computational time, however, as the simulations here remain simple this is not a significant issue. 

Similar to results presented in \cite{Tadday2010CALICEBirks} we found that the value of $kB$ depends strongly on the chosen parameters here. The agreement between simulations and measurements increases with decreasing $\rho$ and was found to stabilize at $0.1\,\mathrm{nm}$. The relation to $\alpha$ was more complex. While a generally good agreement was found for $\alpha=0.1$, it became significantly worse for values of $0.01$ before improving again when going smaller. This is similar to the behavior found in \cite{Tadday2010CALICEBirks} where the agreement converges again when decreasing $\alpha=0.0001$. While the results with $\alpha=0.1$ and $\alpha=0.0001$ were very similar, it was decided for the study here that, although more computationally expensive to use $\alpha=0.0001$ as it is closer to reality. 

The visible energy from the simulations was broadened to take into account the energy resolution of the system. For this, Gaussian broadening was applied to the visible energy. The width of the broadening was taken from the measurement data taken in the 25 to 40~keV energy range as shown in the right of Figure \ref{fig:ADC_POLAR}. The relation of the energy resolution as a function of the beam energy was fitted using a function of the form:

\[\sigma(E) = \sqrt{aE}-b\]

where $\sigma(E)$ is the Gaussian width of the photo-peak, $E$ is the total deposited energy and a and b are free parameters. To calculate the energy broadening for each event in the simulations, the total deposited energy is used in this equation to calculate $\sigma$ and a random number is picked from the Gaussian distribution centered around the visible energy using the width $\sigma$.

By only using the fit in the 25 to 40~keV energy range to produce the energy resolution relation from the data, we fix this to be equal to the measurements in this range. In the case of a simple system with a linear gain it would continue to follow this relation after 40 keV. However, any deviations from the simple $\sqrt{E}$ relation above this energy will be a result of combinations of Birks' law and effects induced by the K-edge.

It should be noted that the energy broadening does not have a significant effect on the measurement of $kB$ discussed in the next section. Rather, it was used to see if a jump in energy resolution is observed in the simulations as well above the K-edge and to perform a qualitative comparison with the measurement results. As the exact details on the energy resolution depend also significantly on the electronics and the SiPM coupling, a quantitative comparison would not be appropriate here.

\section{Analysis}\label{sec:analysis}

\begin{figure}[h!]
\centering
 \includegraphics[width=\textwidth]{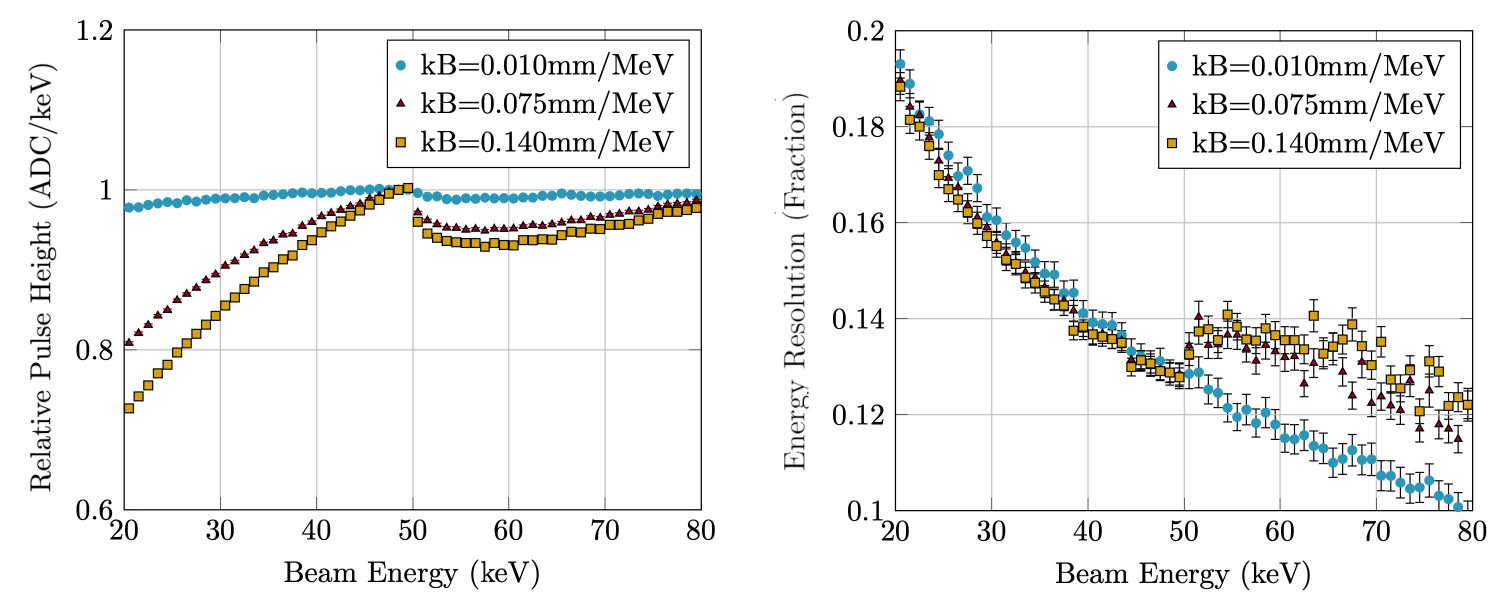}
 \caption{\textbf{Left:} Equal to Figure \ref{fig:ADC_POLAR} but produced using the simulations with $kB=0.010\,\mathrm{mm/MeV}$ (dots), $kB=0.075\,\mathrm{mm/MeV}$ (triangles) and $kB=0.140\,\mathrm{mm/MeV}$ (squares). \textbf{Right:} The energy resolution as a function of energy for the same 3 values of $kB$.} 
 \label{fig:example1}
\end{figure}

Simulations were performed for beam energies ranging from 10.7 keV to 100.7 keV in steps of 1 keV. The addition of the $0.7\,\mathrm{keV}$ allows the simulation results to line up directly with the measurements where data from the HPGe detector indicated an offset of $0.7\,\mathrm{keV}$. Each simulated spectrum, which is Birks corrected and includes energy broadening, was fitted with a Gaussian function where the mean ($\mu$) was taken as the mean pulse height and the $\sigma/\mu$ as the energy resolution. In order for the simulation results of the relative pulse height to appear on the same scale as the measurements, the histograms were normalized to their heights point which is typically at $49.7\,\mathrm{keV}$. These simulations were repeated for different values of $kB$. An example of the simulated pulse height and energy resolution as a function of energy for $kB=0.010\,\mathrm{mm/MeV}$, $kB=0.075\,\mathrm{mm/MeV}$ and $kB=0.140\,\mathrm{mm/MeV}$ can be seen in Figure \ref{fig:example1}. It is clear that with a small Birks' coefficient the relative light yield is nearly flat, while it becomes smaller at low photon energies with increasing $kB$. In addition, the drop above the K-edge becomes more significant as this value increases. In addition, for the energy resolution it can be observed that for low values of $kB$ the energy resolution follows a continuous trend, while the jump which is also observed in the measurement data, becomes clear for  $kB=0.075\,\mathrm{mm/MeV}$. When increasing the value of $kB$, the jump increases slightly in size. 

\begin{figure}[h!]
\centering
 \includegraphics[width=0.7\textwidth]{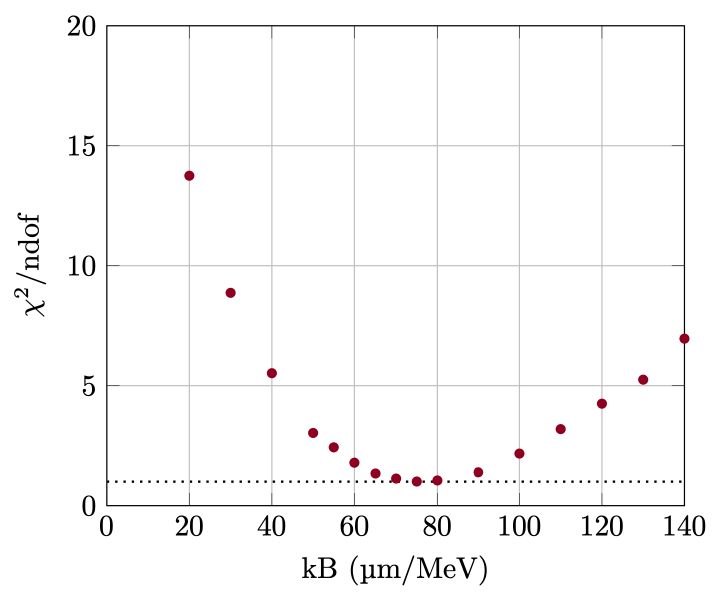}
 \caption{The reduced $\chi^2$ as a function of the simulated value of $kB$. The reduced $\chi^2$ was calculated by fitting the measured relative pulse height as a function of energy against the simulated one.} 
 \label{fig:chi2}
\end{figure}

\begin{figure}[h!]
\centering
 \includegraphics[width=.7\textwidth]{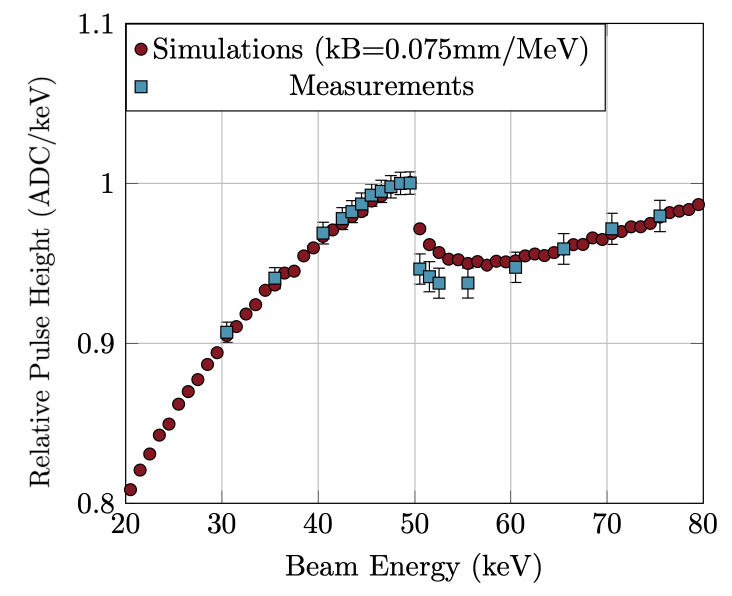}
 \caption{A comparison of the relative pulse height as a function of the beam energy from data (squares) and simulations for $kB=0.075\,\mathrm{mm/MeV}$.}
 \label{fig:best_fit}
\end{figure}

The value of $kB$ was initially varied in steps of $10\,\mathrm{\mu m/MeV}$ in the $10-140\,\mathrm{\mu m/MeV}$ range. For each value a histogram containing the relative pulse height as a function of energy histogram was produced and fitted to the measured data. The goodness of fit was calculated using the reduced $\chi^2$ ($\chi^2$ divided by number of degrees of freedom). The $\chi^2$ values as a function of $kB$ can be seen in Figure \ref{fig:chi2}. This result shows an optimum value around $kB=75\,\mathrm{\mu m/MeV}$. The data and simulated histograms for this value are shown together in Figure \ref{fig:best_fit}. Apart from matching the measured results of the pulse height versus energy well with a $\chi^2/d.o.f.$ of 1.07, this value of $kB$ also results in a relative jump in the energy resolution of $\sim1\%$. 

As the offset of 0.7~keV was only measured for a 50~keV and 100~keV beam energy, we cannot confirm that this offset was exactly 0.7~keV for each beam energy. We therefore repeated the study with random offsets added to the set beam energy picked from a normal distribution centered at 0.7~keV with a $\sigma$ of 0.5~keV. A random offset was picked for each beam energy. Repeating this the analysis 50 times with such random offsets for all beam energies, we consistently found kB to be between $kB=70\,\mathrm{\mu m/MeV}$ and $kB=80\,\mathrm{\mu m/MeV}$. 

\begin{figure}[h!]
\centering
 \includegraphics[width=.7\textwidth]{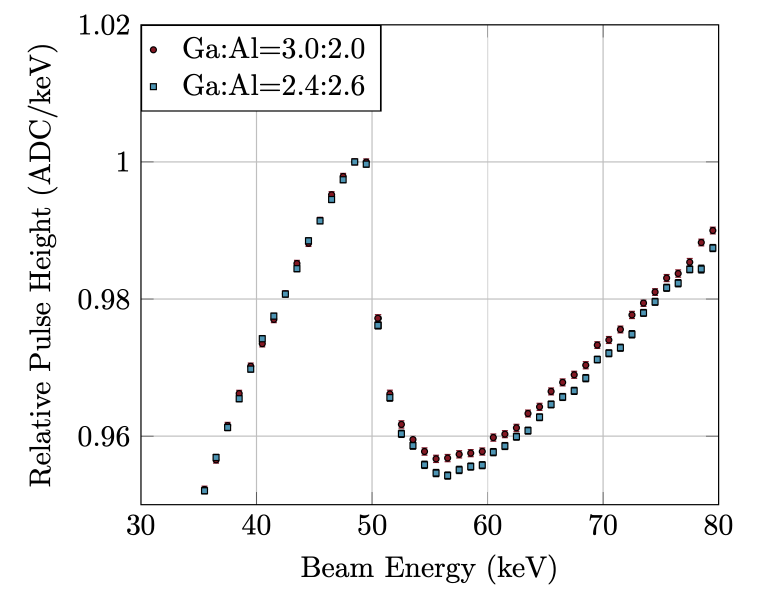}
 \caption{A comparison of the relative pulse height as a function of the beam energy for two simulations with $kB=0.060\,\mathrm{mm/MeV}$ and different gallium to aluminum ratios.}
 \label{fig:ratio}
\end{figure}

It can be observed that the first 4 data points above the K-edge are systematically lower in the measurement data than in the simulations. At higher beam energies the two lines overlap again, while also below the K-edge the two non-linear effects follow one another very closely. While a higher value of $kB$ increases the steepness of the drop above the K-edge it also increases the non-linearity at the energies below it. Increasing the $kB$ will improve the match at higher energies but will therefore result in a worse match at low energies. The sharpness in the drop can, however, also be affected by other properties. One example of this is the Ga:Al ratio in GAGG:Ce, a ratio which can be adjusted to either produce a better energy resolution or higher light yield \cite{GaAl}. For the studies presented here, the standard Ga:Al ratio of 3:2 was used in the simulations. However, an example of the gain to eneryg relation when this is dropped to 2.4:2.6 is compared to the standard ratio in Figure \ref{fig:ratio}. It is clear that reducing the content of Ga, the second-highest Z element in the crystal, will increase the steepness of the drop. This is a result of a higher probability for interactions to take place in the Gd when reducing the amount of Ga. Therefore the Gd K-edge absorption becomes more probable. The behavior below the K-edge does, as expected, remain the same.

A value for the Ga:Al ratio of 2.4:2.6 remains within values found in the literature \cite{GaAl}. Although using this value results in an overall better agreement with the measured data, the drop in relative pulse height remains larger than in the simulations. Re-performing the analysis with this ratio results in a best fit for $kB=72\,\mathrm{\mu m/MeV}$ with a reduced $\chi^2$ of 1.01.

\section{Conclusion and Discussion}

A GAGG:Ce crystal used on the POLAR-2 spectrometer prototype was irradiated at the LARIX-A X-ray facility with mono-energetic X-ray beams in the 30 to 80 keV energy range. The measured data shows a clear drop in the pulse height around the K-edge along with a jump in the energy resolution at the same energies. These measurements can be explained through the effects of Birks' law. The value of  Birks' coefficient $kB$ for GAGG:Ce was derived by comparing our measurements to simulations in which $kB$ is varied. Our results find that the measurements using the GAGG:Ce crystals used in the POLAR-2 prototype match best with a $kB=0.075\,\mathrm{mm/MeV}$. This is significantly higher than most other inorganic scintillators. Something which was also found in \citep{Furuno:2021cns} who report a value of $0.0065\,\mathrm{(g/mm^2)/MeV}$ or $kB=0.039\,\mathrm{mm/MeV}$ using proton and alpha irradiation data. 

Although the results here are not identical to those found in \citep{Furuno:2021cns} it is important to note that the value of $kB$ found can differ significantly based on the analysis performed \cite{BGO}. In part this is due to Birks' law being an empirical formula therefore not describing the data perfectly. The difference between the data and the predictions from the equation will differ depending on the energy scale and type of ionizing radiation used, thereby resulting in variations in the acquired value of kB. As a result, the value found here will allow to accurately describe the behavior for irradiation by hard X-rays, while those found in \citep{Furuno:2021cns} will, likely, provide a more accurate description for high energy ions. In addition, differences in non-linear behavior for different GAGG:Ce crystals analyzed with the same method have been reported \cite{campana_etal_2023}. The data reported there, taken at the same facility, can be used to derive the $kB$ as well. The authors there present 3 sets of 2 non-linear functions which describe the gain around the K-edge from their 3 GAGG:Ce crystals. One of these functions describes the gain below the K-edge while the other describes that above it. We can fit this data using the same simulation results used for our study and find respective values of $kB=0.048\,\mathrm{mm/MeV}$, $kB=0.055\,\mathrm{mm/MeV}$ and $kB=0.060\,\mathrm{mm/MeV}$. To derive these values, the simulations were fitted against the empirical functions rather than the measured data points, therefore not allowing for a very accurate measurement. 

It should finally be noted that our best fit for $kB$ changes when the Ga:Al ratio in the GAGG:Ce is changed. As we do not know this ratio for our particular material, we performed our analysis using the most extreme ratio found in the literature. Using this ratio results in a better agreement with our data with a lower reduced $\chi^2$. Potentially indicating that the type of GAGG:Ce used here has a lower Ga content. We therefore encourage future measurements of this type with crystals with known Al:Ge ratios.

\section*{Acknowledgements}
We gratefully acknowledge Lisa Ferro from the University of Ferrara for operating the beam and assisting during the LARIX-A calibration campaign. This work is supported by the AHEAD-2020 Project grant agreement 871158 of the European Union’s Horizon 2020 Program, which made the calibration with the GAGG crystal possible. 

\appendix

\bibliographystyle{elsarticle-num} 


\end{document}